\documentclass[9pt,twocolumn,twoside]{osajnl}
	\usepackage{flushend}
	
	\journal{ao} 
	
	\setboolean{shortarticle}{false} 
	
\title{Design considerations of high-performance InGaAs/InP single-photon avalanche diodes for quantum key distribution}
	
\author[1,2]{Jian Ma}
\author[1,2]{Bing Bai}
\author[1,2]{Liu-Jun Wang}
\author[3]{Cun-Zhu Tong}
\author[1,2]{Ge Jin}
\author[1,2,*]{Jun Zhang}
\author[1,2]{Jian-Wei Pan}
\affil[1]{Hefei National Laboratory for Physical Sciences at the Microscale and Department
of Modern Physics, University of Science and Technology of China, Hefei, Anhui 230026, China}
\affil[2]{CAS Center for Excellence and Synergetic Innovation Center in Quantum Information
and Quantum Physics, University of Science and Technology of China, Hefei, Anhui 230026, China}
\affil[3]{State Key laboratory of Luminescence and Application, Changchun Institute of Optics, Fine Mechanics and Physics, Chinese Academy of Sciences, Changchun 130033, China}

\affil[*]{Corresponding author: zhangjun@ustc.edu.cn}
	
	\dates{Compiled \today}
	
\ociscodes{(040.1345) Avalanche photodiodes (APDs); (040.5570) Quantum detectors; (270.5568) Quantum cryptography}
	
	\doi{\url{http://dx.doi.org/10.1364/ao.XX.XXXXXX}}
	
	\begin{abstract}
InGaAs/InP single-photon avalanche diodes (SPADs) are widely used in practical applications requiring near-infrared photon counting such as quantum key distribution (QKD).
Photon detection efficiency and dark count rate are the intrinsic parameters of InGaAs/InP SPADs, due to the fact that their performances cannot be improved using different quenching electronics given the same operation conditions. After modeling these parameters and developing a simulation platform for InGaAs/InP SPADs, we investigate the semiconductor structure design and optimization. The parameters of photon detection efficiency and dark count rate highly depend on the variables of absorption layer thickness, multiplication layer thickness, excess bias voltage and temperature. By evaluating the decoy-state QKD performance, the variables for SPAD design and operation can be globally optimized. Such optimization from the perspective of specific applications can provide an effective approach to design high-performance InGaAs/InP SPADs.
	\end{abstract}
	
	\setboolean{displaycopyright}{true}
	
\begin{document}
	
	\maketitle
	\thispagestyle{fancy}
	\ifthenelse{\boolean{shortarticle}}{\abscontent}{}
	
	\section{Introduction}
Single-photon detection~\cite{N09,EFM11} in the near-infrared range is a key technique for many areas such as quantum key distribution (QKD) and Lidar. For practical applications, using III-V compound semiconductor devices, e.g., InGaAs/InP single-photon avalanche diodes (SPADs), is a currently mainstreaming solution due to the advantages of low cost and small size~\cite{JMO11,ZIZ15}. The key parts of an InGaAs/InP single-photon detector include a SPAD device~\cite{JMO07} and quenching electronics~\cite{AO96}. In the early stage, commercial avalanche photodiodes for fiber-optic communication were often exploited to be operated in Geiger mode~\cite{LZC96,RGZ98,RWR00}. However, since such devices were initially designed for linear mode operations, their performance for Geiger mode operations was poor. Therefore, designing optimized devices dedicated to near-infrared single-photon detection is crucial, and so far various groups have been working in this direction~\cite{JMO07,HBL00,JQE06,DDM06,PJ13,OE14}.

For InGaAs/InP SPADs, there are diverse parameters to characterize its performance including photon detection efficiency (PDE), dark count rate (DCR), afterpulse probability, timing jitter, maximum count rate~\cite{ZIZ15}. Normally, InGaAs/InP SPADs are designed for general purposes, and thus all the SPAD parameters have to be compromised during the design process. With the considerations from the perspective of applications, one may focus on the key parameters to further optimize the design of SPADs, which helps to fabricate application-specific devices and brings performance improvements for applications.

For instance, among the above parameters of InGaAs/InP SPADs, using the technique of high-frequency gating~\cite{NSI06,Toshiba07,GAP09,NTY11,LLW12,RBM13} afterpulse probability can be significantly suppressed and thus maximum count rate can be greatly increased, which is well suited for QKD applications~\cite{ZIZ15}. Therefore,
the most concerning and intrinsic parameters when designing InGaAs/InP SPADs for QKD are PDE and DCR, which are independent from quenching electronics~\cite{ZIZ15}. This means that the performance of PDE and DCR cannot be improved using different quenching electronics under the same operation conditions.

In this paper, after modeling PDE and DCR for InGaAs/InP SPADs, we develop a simulator for SPAD design and focus on the device structure optimization to obtain better performance for QKD applications. This application-driven optimization provides a new approach for SPAD structure design, and the simulation result could be an important reference for the SPAD fabrication in practice.

\section{Device structure and modeling}

\begin{figure}[tpb]
  \includegraphics[width=8.2cm]{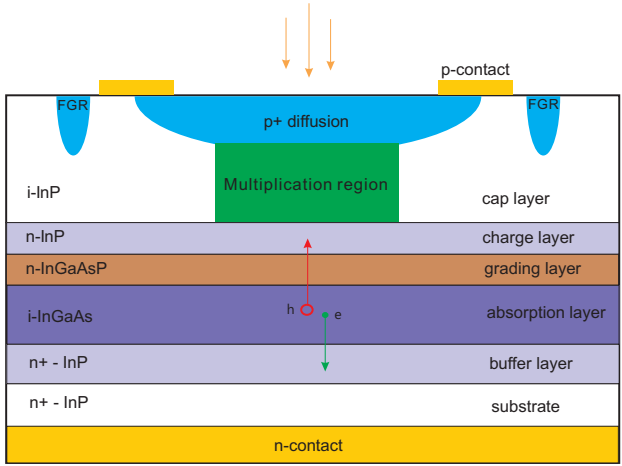}
  \caption{Structure of InGaAs/InP SPAD}
  \label{fig1}
\end{figure}

The typical heterojunction of InGaAs/InP SPADs are based on separate absorption, grading, charge, and multiplication (SAGCM) structure~\cite{JMO07}, as shown as Fig \ref{fig1}.
With reverse bias, such structure can guarantee high electric field in the multiplication layer to enhance the avalanche probability while relatively low electric field
in the absorption layer to reduce dark current due to tunneling effects~\cite{JMO07}. Meanwhile, the grading layer is inserted between the absorption and multiplication layers to avoid the accumulation of trapped carriers during avalanche process so that the afterpulsing effect can be reduced, and the
charge layer between the absorption and multiplication layers is to tailor the internal electric field profile~\cite{JMO07}.

Given an InGaAs/InP SPAD operated in Geiger mode, the reverse bias is above the breakdown voltage. In the absorption layer, an incoming photon may generate an electron-hole pair. Due to the electric field, the hole passes across the InGaAsP grading layer and the InP charge layer, and is finally drifted to the multiplication layer, in which a self-sustaining avalanche may be created. The avalanche is then sensed by a discriminator and quenched with a readout circuit by lowering the bias below the breakdown voltage.

The parameter of PDE~\cite{PDE} is defined as the probability of producing a correct output signal in response to an incident photon. Therefore, PDE
of InGaAs/InP SPADs can be expressed by~\cite{ZIZ15}
\begin{equation}
\label{equ1}
PDE=P_{c}P_{abs}P_{inj}P_{ava},
\end{equation}
where $P_c$ is the coupling efficiency, $P_{abs}$ is the absorption efficiency in the absorption layer, $P_{inj}$ is the efficiency of photo-generated carriers injected from the absorption layer to the multiplication layer, and $P_{ava}$ is the avalanche efficiency in the multiplication layer.
For normal pigtailed SPAD devices, $P_c$ is a fixed value. Meanwhile, for SAGCM SPADs after an electron-hole pair is generated the effects of both the recombination in the absorption layer and the accumulation in the grading layer are negligible. In the simulation, $P_c$ and $P_{inj}$ are supposed to be 1 for simplicity.
$P_{abs}$ can be calculated by
\begin{equation}\label{equ2}
 P_{abs}=1-e^{-\alpha\cdot L_{abs}},
\end{equation}
where $\alpha $ is the absorption coefficient of In$_{0.53}$Ga$_{0.47}$As and $L_{abs}$ is the absorption layer thickness.

$P_{ava}$ is defined as the probability of creating a self-sustaining avalanche due to a hole entering into the multiplication layer, which highly depends on the electronic field or excess bias and temperature. In the model, the ionization coefficients of electron and hole are taken from the reference by Zappa \emph{et al.}~\cite{ZLL96}, which have been used to fit experimental results previously~\cite{DDM06}. The electric field and temperature dependence of $P_{ava}$ is then calculated using the model proposed by Oldham \emph{et al.}~\cite{OSA72}, in which the dead space effect~\cite{OC72} have not been considered. As discussed in the references~\cite{HST92,M99}, for SPADs with thick multiplication layers the dead space effect on carrier multiplication is negligible. Since the multiplication layer thickness of designed SPAD is normally larger than 1 $\mu$m, using the model of Oldham \emph{et al.}~\cite{OSA72} to calculate $P_{ava}$ is appropriate.

For the parameter of DCR~\cite{DCR}, three generation mechanisms are considered~\cite{DDM06}, i.e., the thermal generation in the absorption layer, direct band-to-band tunneling and trap assisted tunneling (TAT) in the multiplication layer. The thermal generation induced DCR is primarily due to the Shockley-Read-Hall process and can be calculated as
\begin{equation}
\label{equ3}
DCR_{abs}=\frac{n_i}{\tau_{abs}},
\end{equation}
where $n_i$ is the intrinsic carrier concentration and $\tau_{abs}$ is effective lifetime of free carriers in the In$_{0.53}$Ga$_{0.47}$As layer. In the simulations, $\tau_{abs}$ is assumed to be 50 $\mu$s.

\begin{figure}[tpb]
  \includegraphics[width=8.2cm]{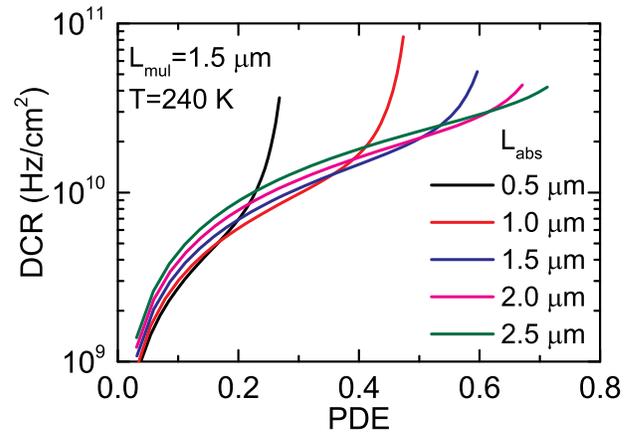}
  \caption{DCR versus PDE with the absorption layer thicknesses ranging from 0.5 $\mu$m to 2.5 $\mu$m. The temperature is 240 K and the multiplication layer thickness is 1.5 $\mu$m.}
  \label{fig2}
\end{figure}

The direct band-to-band tunneling induced dark current density can be calculated as~\cite{DDM06}
\begin{equation}
\label{equ4}
J_{tun\_dir}=AF(x)^2 exp(\frac{-BE_g^{3/2}}{F(x)}),
\end{equation}
where $E_g$ is the bandgap of the multiplication layer, $A=q^3(2m_r/(q E_g))^{1/2}/(4\pi^3\hbar^2)$, $B=\pi(m_r/2)^{1/2}/(2q\hbar)$ and $F(x)$ is the electric field. $q$ is the electric charge and $\hbar$ is the reduced Planck constant. $m_r=m_cm_{lh}/(m_c+m_{lh})$, where $m_c$ and $m_{lh}$ are the effective masses of electrons and light holes, respectively.

During the process of TAT, a carrier is tunneled from the valence band to midgap between the conduction band and the valence band due to a defect, and the trapped carrier is then tunneled to the conduction band. TAT depends on the electric field, temperature, defect density and the activation energy of trapped state. The current density contribution of TAT
can be calculated as~\cite{DDM06}
\begin{equation}
\label{equ5}
J_{tun\_trap}(x)=\frac{AF(x)^2N_Texp(\frac{-(B_1E_{B1}^{3/2}+B_2E_{B2}^{3/2})}{F(x)})}{N_v exp(\frac{-(B_1E_{B1}^{3/2})}{F(x)})+N_c exp(\frac{-(B_2E_{B2}^{3/2})}{F(x)})},
\end{equation}
where $N_v$ and $N_c$ are the densities of states in the valence and conduction bands, $N_T$ is the defect density in the multiplication layer, $B_1=\pi(m_{lh}/2)^{1/2}/(2q\hbar)$, $B_2=\pi(m_c/2)^{1/2}/(2q\hbar)$, and $E_{B1}$ and $E_{B2}$ represent the energy barriers tunneling from the valence band to trap and tunneling from trap to the conduction band, respectively. $E_{B1}=aEg$ and $E_{B2}=(1-a)Eg$, in which $a$ is a fitting parameter~\cite{DDM06}. In the model, $a$ and $N_T$ are assumed to be fixed as 0.75 and $4\times10^{14}cm^{-3}$, respectively, for simplicity of simulations.

In the above model, the electric field distribution $F(x)$ is a crucial parameter, which can be calculated using Gauss’s law given a reverse bias value. The SPAD parameters of breakdown voltage, $P_{ava}$ and DCR highly depend on $F(x)$. Our simulation model is one-dimensional, and pre-mature edge-breakdown effect at the curved interface of the diffusion region has not been considered due to the fact that such effect can be well avoided by double diffusion process or float guard ring process~\cite{floatgr}. Since our model is the same as that in Ref.~\cite{DDM06}, in which simulation results and experimental data agree with each other very well, we compare our simulation results of avalanche probability
distribution with those in Ref.~\cite{DDM06} given the same parameters. The exactly same results in two cases can well validate our simulation model.

\section{Simulation results and discussion}

\begin{figure}[tbp]
  \includegraphics[width=8.2cm]{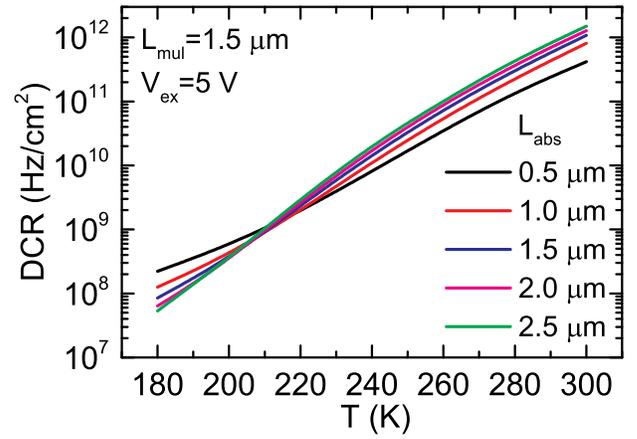}
  \caption{DCR as a function of temperature with the absorption layer thicknesses ranging from 0.5 $\mu$m to 2.5 $\mu$m. $V_{ex}$ is 5 V and the multiplication layer thickness is 1.5 $\mu$m.}
  \label{fig3}
\end{figure}

In this section, we study the tradeoff between DCR and PDE under different conditions using the simulator, and then evaluate the performance in QKD applications. DCR and PDE are highly related to two structure variables, i.e., absorption layer thickness ($L_{abs}$) and multiplication layer thickness ($L_{mul}$), and two operation variables as well, i.e., excess bias voltage ($V_{ex}$) and temperature ($T$).

First, the variables of $L_{mul}$ and $T$ are fixed whilst the variables of $L_{abs}$ and $V_{ex}$ vary, and the results are shown in Fig.~\ref{fig2}. Linearly increasing $V_{ex}$ corresponds to roughly linear increase of PDE and exponential increase of normalized DCR, so that DCR is an exponential function of PDE. When $V_{ex}$ is low, the slope of exponential increase is moderate but becomes pretty steep in the region of high $V_{ex}$, as shown in Fig.~\ref{fig2}. $L_{abs}$ is chosen from 0.5 $\mu$m to 2.5 $\mu$m with a step of 0.5 $\mu$m, and the maximum achievable PDE increases with an increase of $L_{abs}$.

\begin{figure}[bpt]
  \includegraphics[width=8.2cm]{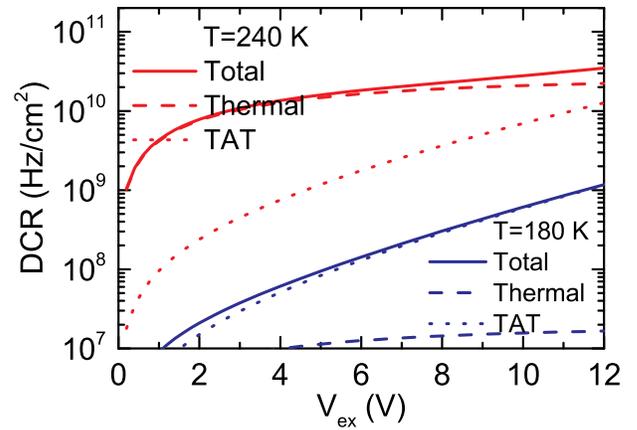}
  \caption{DCR (solid line) as a function of excess bias including the contributions of thermal generation (dashed line) and TAT (dotted line) at 240 K (red) and 180 K (blue). The thicknesses of the absorption and multiplication layers are 1.8 $\mu$m and 1.5 $\mu$m, respectively.}
  \label{fig4}
\end{figure}

\begin{figure}[htbp]
  \includegraphics[width=8.2cm]{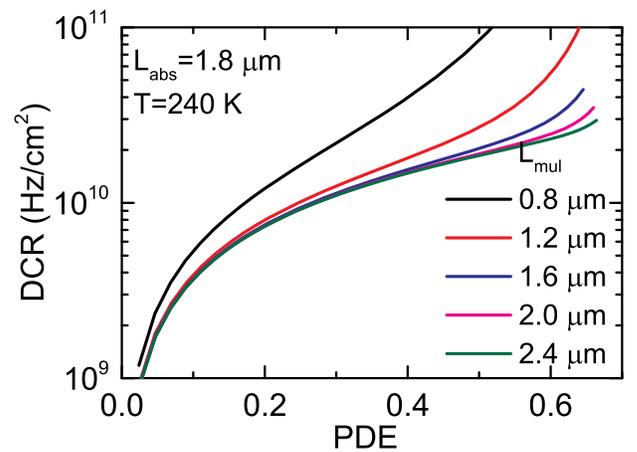}
  \caption{DCR versus PDE with the multiplication layer thicknesses ranging from 0.8 $\mu$m to 2.4 $\mu$m. The temperature is 240 K and the absorption layer thickness is 1.8 $\mu$m.}
  \label{fig5}
\end{figure}

Second, we simulate the characteristic of temperature dependence. Fig.~\ref{fig3} shows DCR as a function of temperature in the cases of
different absorption layer thicknesses, given fixed $L_{mul}$ of 1.5 $\mu$m and $V_{ex}$ of 5 V. With a decrease of temperature,
DCR decreases exponentially, and from Fig.~\ref{fig3} one can observe DCR reduction with 4$\sim$5 orders of magnitude when the temperature is cooled down from 300 K to 180 K. Such reduction is primarily due to the contribution of thermal generation, since thermal generation depends on $n_i$ from Eq.~\ref{equ3} and further $n_i$ decreases drastically with a decrease of temperature~\cite{PRB}. Moreover, one may find out an interesting phenomenon in Fig.~\ref{fig3}, i.e., at temperatures higher than around 210 K larger $L_{abs}$ induces higher DCR but this trend is reverse at temperatures below 210 K. This is probably due to the fact that thermal generation and TAT dominate DCR at high and low temperatures, respectively~\cite{JMO07}. Under the conditions as shown in Fig.~\ref{fig3}, 210 K may be the temperature turning point between the two DCR generation mechanisms.

Third, to verify the contribution difference between the two DCR mechanisms, DCRs as a function of $V_{ex}$ at two typical temperatures, i.e., 240 K in the high temperature region and 180 K in the low temperature region, are compared as shown in Fig.~\ref{fig4}. Clearly, at 240 K thermal generation dominates DCR and the increase slope of DCR is flat particularly in the region of large $V_{ex}$. Similarly, at 180 K TAT dominates DCR and the increase slope of DCR becomes much steeper than that at 240 K.

\begin{figure}[htbp]
  \includegraphics[width=8.2cm]{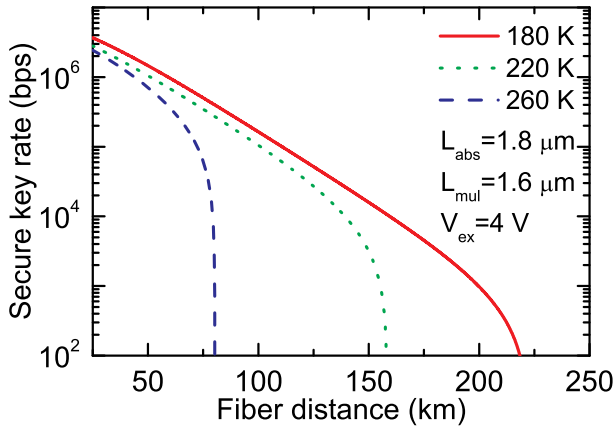}
  \caption {Secure key rate as a function of fiber distance at 180 K (red solid line), 220 K (green dotted line) and 260 K (blue dashed line). The thicknesses of the absorption and multiplication layers are 1.8 $\mu$m and 1.6 $\mu$m, respectively.}
  \label{fig6}

\end{figure}

Finally, the variables of $L_{abs}$ and $T$ are fixed whilst the variables of $L_{mul}$ and $V_{ex}$ vary. Fig.~\ref{fig5} exhibits DCR as a function of PDE with $L_{mul}$ ranging from 0.8 $\mu$m to 2.4 $\mu$m. Given a PDE value, larger $L_{mul}$ results in lower DCR performance. However, larger $L_{mul}$ also induces worse timing jitter and higher afterpulse probability~\cite{JMO11}.

Therefore, to obtain good DCR and PDE performances the four variables have to be optimized. We propose an approach by evaluating the QKD performance to optimize SPAD parameters. In fiber-based QKD applications, the most important parameters are fiber distance and secure bit rate. Through QKD performance evaluation, one may globally optimize the structure and operation variables for SPADs.
In the QKD simulations, decoy-state BB84 protocol~\cite{Wang05,Lo05,Wang05pra,Ma05} using polarization encoding is used. The parameters and the corresponding values are listed in Table~\ref{table:QKD}, some of which are taken from the reference~\cite{Liu10}. Since the afterpulse probability ($P_{ap}$) of SPAD is highly related with the quenching electronics~\cite{ZIZ15}, for simplicity it is assumed that the values of $P_{ap}$ are fixed at 0.01 and 0.03 in the regions of low and high $V_{ex}$.
For the SPADs used in the QKD simulations, the diameter of active area is set as 25 $\mu$m, and the parameters of $L_{abs}$ and $L_{mul}$ are fixed at 1.8 $\mu$m and 1.6 $\mu$m, respectively.

\begin{table}[t]
\caption{Parameters used in the QKD simulation.} 
\centering 
\begin{tabular}{l l} 

\hline 
Parameter & Value \\
\hline
mean photon number of signal states ($\mu$)  &  0.6 \\
\hline
mean photon number of decoy states ($\nu$) & 0.2   \\
\hline
probability ratio of the 3 intensities & 6:1:1 \\
\hline
repetition frequency ($f$) & 1 GHz \\
\hline
error correction factor ($F$) & 1.2 \\
\hline
optical fiber attenuation ($\alpha_f$) & 0.2 dB/km \\
\hline
insertion loss in receiver ($t$) & 3 dB \\
\hline
error rate due to optical imperfections ($e_o$) & 0.01 \\
\hline
afterpulse probability of SPAD ($P_{ap}$) & 0.01 ($V_{ex}\leq 4 V$); \\
          &                                 0.03 ($V_{ex}> 4 V$) \\

\hline 
\end{tabular}
\label{table:QKD} 
\end{table}

\begin{figure}[hbtp]
 \centering
  \includegraphics[width=8 cm]{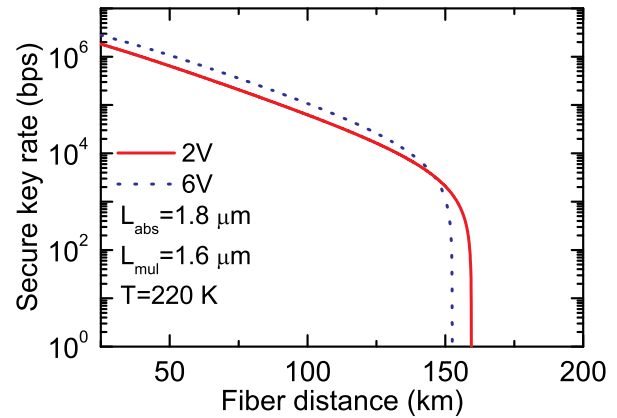}\\
  \caption{Secure key rate as a function of fiber distance with the excess bias voltages of 2 V (red solid line) and 6 V (blue dotted line) at 220 K. The thicknesses of the absorption and multiplication layers are 1.8 $\mu$m and 1.6 $\mu$m, respectively.}
  \label{fig7}
\end{figure}

Given a fixed $V_{ex}$, the simulation results at different temperatures are shown in Fig.~\ref{fig6}. In the region of short distances, the secure key rates are pretty high, and the key rate differences in three cases are small. Moreover, as temperatures decrease the maximum distances significantly increase. In the region of long distances, due to the low photon flux, DCR may dominate the quantum bit error rate so that DCR has to be greatly reduced in order to increase the secure key rate and distance. This also indicates that cooling down the SPAD temperature can be an effective method in order to further boost the maximum QKD distance, which has been experimentally demonstrated to achieve a record QKD distance of 307 km using an ultra-low noise InGaAs/InP SPAD~\cite{KLH15}.

Similarly, Fig.~\ref{fig7} shows the simulations with different $V_{ex}$ at the same temperature. Larger $V_{ex}$, corresponding to higher PDE, DCR and $P_{ap}$, has higher secure key rate and shorter maximum distance. Combining the results in Fig.~\ref{fig6} and Fig.~\ref{fig7}, one can conclude that $V_{ex}$ and temperature are the variables to be considered in prior for the aims of high key rate and long distance, respectively. Given a fixed QKD distance, through the optimization of $V_{ex}$ and temperature the secure key rate can be maximized. Further, the variables of $L_{abs}$ and $L_{mul}$ can be optimized using better QKD performance as a reference, i.e., higher key rate and longer distance, under the conditions of the same $V_{ex}$ and temperature.

\section{CONCLUSIONS}

In summary, we have modeled and developed an integrated simulation platform for InGaAs/InP SPADs. We focus on the intrinsic parameters of PDE and DCR, whose performance dependence on $L_{abs}$, $L_{mul}$, $V_{ex}$ and temperature has been evaluated. To globally optimize the four variables for SPAD design and operations, the parameters of PDE and DCR are applied in the decoy-state QKD simulations, and the QKD performance is an important reference to determine the values of variables. This approach could be effective for designing high-performance InGaAs/InP SPADs in practice.
	
\section*{Acknowledgments}
This work has been financially supported by the National Basic Research Program of China Grant No.~2013CB336800, the National Natural Science Foundation of China Grant No.~61275121, and the Chinese Academy of Sciences.

	\end{document}